\documentclass[a4,4pt]{article}
\usepackage{amssymb,calligra,
mathrsfs}
\usepackage{amsmath}
cm \textheight 22 true cm \topmargin=-2mm

\begin{document}
\thispagestyle{empty}\setcounter{page}{1}
 \vskip40pt
\centerline{\bf Co-operativity  in neurons and the role of noise
in  brain} \centerline{\bf }

\vskip10pt

\centerline{\footnotesize Indranil Mitra$^1$, Sisir Roy$^{2,3}$, Gary Hastings$^1$}
\vskip5pt \centerline{\footnotesize $^{1}$ Brain \&\ Behaviour Program,
Department of Physics, Georgia State University}
\centerline{\footnotesize 29 Peachtree Center Avenue, Atlanta,GA
30303 USA } \vskip5pt \centerline{\footnotesize$^{2}$Physics and
Applied Mathematics Unit, Indian Statistical Institute,}
\centerline{\footnotesize 203 Barrackpore Trunk Road, Kolkata
700108, India}

\centerline{\footnotesize$^{3}$College of Science,
George Mason University} \centerline{\footnotesize 4400 University
Drive Fairfax, Virginia 22030,USA}

\centerline{\footnotesize$^{1}$e-mail: imitra@gsu.edu }
\centerline{\footnotesize$^{2}$e-mail: sisir@isical.ac.in}
\centerline{\footnotesize$^{1}$e-mail: ghastings@gsu.edu}

\vskip10pt

\abstract

In view of some recent results  in case of the dopaminergic
neurons exhibiting long range correlations in VTA of the limbic
brain we are interested to find out whether any stochastic
nonlinear response may be reproducible in the nano scales usimg
the results of quantum mechanics. We have developed a scheme  to
investigate this situation in this paper by taking into
consideration the Schrodinger equation (SE) in an arbitrary
manifold with a metric, which is in some sense a special case of
the heat kernel equation. The special case of this heat kernel
equation is the diffusion equation, which may reproduce some key
phenomena of the neural activities. We make a dual equivalent
circuit model of SE  and incorporate non commutativity and noise
inside the circuit scheme. The behaviour of the circuit elements
with  interesting limits are investigated. The most bizarre part
is the long range response of the model by dint of the Central
Limit Theorem, which is responsible for coherent behaviour of a
large assembly of  neurons.

\newpage
\newpage

\section{Introduction}

Nervous systems use electrical signals which propagate through ion
channels which are specialized proteins and provide a selective
conduction pathway, through which appropriate ions are escorted to
the cell's outer membrane. Also, the ion channels undergo fast
conformational changes in response \cite{hill} to metabolic
activities which opens or closes the channels as gates. The gating
essentially involves changes in voltages across the membrane and
ligands. The voltage dependent ion channels have an ability to
alter ion permeability of membranes in response to changes in
transmembrane potentials. The channels which are Na, K and Ca
voltage gated or synaptic channels gated by acetylcholine,
glycine or $\gamma$ aminobutyric acid seemed similar. The
magnitude of current across membrane depends on the density of
channels, conductance of the open channel and how often the
channel spends in the open position or the probability. Hodgkin
and Huxley \cite{hh1,hh01} accounted for the voltage sensitivity
of Na$^{+}$ and K$^{+}$conductance of the squid giant axon by
postulating charge movement between kinetically distinct states
of hypothetical activating particles. In spite of the detail
electrophysiological studies, the atomic structure of voltage
gated ion channels still remained in the dark till the discovery
of Mckinnon and his collaborators \cite{mck00,mck,mck1} which
obtained a crystal structure of a $Ca^{2+}$ gated $K^{+}$ ion
channel provides a mechanism for gating \cite{gate,gate2}.A
functional study of $KvAP$ in this context led to a proposal
known as the voltage sensor paddle model. Ion channels are
membrane-spanning proteins with central pores through which ions
cross neuronal membranes.
 The pores through each ion channel flicker between open and closed states, starting and stopping the flow of ions and the
 electrical current they carry.

Considering the voltage sensor capabilities of the ion channels
and generation of currents and potentials, we in this paper deal
mainly with the electrical properties of the ion channels. It is
already known that the neuron acts as an electrical device,
\cite{nelec} where a potential difference develops across the
membrane due to differences in ion concentrations between inside
and outside the cell. The participating ions are
Sodium($Na^{+}$), Potassium $(K^{+}$), Calcium ($Ca^{+}$) and
Chlorine($Cl^{+}$). Nernst equation describes equilibrium
potential for a single ionic species as $E =
\frac{RT}{zF}ln\frac{[X^{+}]_{o}}{[X^{+}]_{i}}$. Total membrane
current is given by the sum of individual channel currents
$$I_{m} = I_{Na} + I_{K} + I_{Cl}$$

In this way, a membrane patch can be described by an equivalent
electrical circuit component. As we have discussed earlier,
electrical signals are changed in the membrane potential at
specific sites of the neuronal network, which are obtained by
changes due the closing and opening of ion channels. Given these
things to be known, the main objective of this paper is
different. First of all in a recent article \cite{qmic} it has
been hypothesized from dimensional arguments that quantum
mechanics may be operative at some scale in the ion channels. If
this is the case then the whole story of voltage sensing in ion
channel gets a new paradigm shift. If we assume that membrane
voltage and currents are generated through equivalent circuits
but at length scales where quantum mechanics is assumed to hold,
then due to noncommutative effects the whole concept of devising
electrical circuits is different, but also at the same time it
should be mentioned here that at large length scales
corresponding to a large collection of ion channels in comparison
to a single or few ion channels in the previous case, we expect
the quantum effects to average out and the conventional circuit
elements for describing the mechanisms of voltage
\cite{walk,nelec1} and current generation through the gates is
valid.

In this paper, we implement a quantum circuit for the ion channels
following the lines of \cite{qcio2}. The fundamental assumption
has been that, Hodgkin and Huxley's empirical, deterministic model
may be reformulated where the relevant behaviour arises from the
combined contributions of a large number of small stochastic
components. Our model averages away the random behaviour at the
smallest scale.Our intention in this paper is to set up
rigorously the underlying stochastic process of one such model
and then to prove that its behaviour converges to that predicted
by the earlier deterministic model as the limit is taken in a
suitable regime, in that the sample paths of the stochastic
process converge in probability to the trajectory predicted by the
deterministic model. Stochastic Hodgkin-Huxley model relates
behaviour on three distinct scales, the flow of charge at the
scale of individual ions, the opening and shutting of ion
channels at the scale of large protein molecules and the working
of the whole axon. The basic task in hand is as follows, we have
developed a Schrodinger equation and a implementation of the
equivalent circuit. Now following the work \cite{amaq} on
neuromanifolds we assume that the underlying geometry of the ion
channels is not known a priori. Thereby we assume a curved
manifold and write down the nonlinear Schr\"odinger equation
(NLSE), essentially a heat kernel equation in a curved manifold
\cite{nlse00}. The next task in hand is to find out an equivalent
circuit model for that. In the last section we find out a
connection with the HH model and determine how the quantum
effects may get lost at large length scales in the mesoscopic
case when we take the limiting case of large number of ion
channels. Thus we can think of the difference between the
stochastic and deterministic models as one of resolution:
although neither model can `see' the individual ions, the
stochastic model can see single channels, whereas even these are
beyond the deterministic model. In this sense the main result of
this paper is a check that if we average out over the smallest
scale to obtain the stochastic model, and then consider a
suitable limit of this to represent the vanishing size of the
intermediate scale, we recover the model obtained by averaging
over both smallest and intermediate scales from the start.So
essentially in this paper we are confronted in understanding the
dependence of noise and co-operative effect of the neuronal
architecture on the spatial and temporal scales in the brain.

\section{Role of Noise \&\ Cooperativity in Hodgekin-Huxley(HH) Formalism}

Following the study of Hodgkin and Huxley, most of the models of axons have treated the generation and propagation of action potentials using deterministic differential equations. Since \cite{lecnoss} it has become increasingly evident, however, that not only the synaptic noise but also the randomness of the ion channel gating itself may cause threshold fluctuations in neurons \cite{fluc1,fluc2}. Therefore, channel noise which originates in the stochastic nature of the ion channel dynamics should be taken into account \cite{iod1}. For example, in mammalian ganglion cells both the synaptic noise and the channel noise might equally contribute to the neuronal spikes variability \cite{neurspik}. Due to a finite size, the origin of the channel noise is basically due to fluctuations of the mean number of open ion channels around the corresponding mean values. Therefore, the strength of the channel noise is mainly determined by the number of ion channels participating in the generation of action potentials. Channel noise impacts, for example, such features as the threshold to spiking and the spiking rate itself the anomalous noise-assisted enhancement of transduction of external signals, i.e. the phenomenon of stochastic resonance \cite{stochr1,stochr2,stochr3}, and the efficiency for synchronization.

When an ion channel opens or closes, an effective gating charge is moved across the membrane. This motion creates the so-called gating current which is experimentally measurable \cite{exp1}. The influence of gating currents was not explicitly considered in the original Hodgkin–Huxley (HH) model. The model we would like to describe is a neuronal model at length
scales of ion channels where we believe that quantum mechanics may
be operative. But we believe that the model may also include the HH
model as a special case where coarse graining can be done, or for
example, if we include large number of channels, the collective
behaviour should be described by the HH model. For the sake of
completeness, we would like to describe the HH model in brief
\cite{hh2}.

In the HH case, the basic membrane circuit suitable for, say, a
squid giant axon with two voltage dependent channels is given by the
following construction:
 The circuit is described by a capacitor $C$,
sodium, potassium, leakage conductance $G_{Na}$, $ G_{K}$ and
$G_{L}$ respectively. The membrane potential is the voltage
difference between the outside and inside of the cell membrane and
there can be a current injected into the cell from an electrode or
other parts of the cell.

 The equations describing the phenomena is given by
\begin{equation}\label{hh111}
C \frac{dV}{dt}= I_{ext} - G_{Na}(V-E_{Na}) - G_{K}(V-E_{K}) -
G_{L}(V-E_{L})
\end{equation}

$G_{Na}$ and $G_{K}$ are the functions of membrane potentials and
time and are given by the following equations,
\begin{eqnarray}\label{hh222}
G_{Na} &=& \overline{G_{Na}}m^{3}h \quad \frac{dm}{dt} =
\frac{m_{\infty}(V) - m }{\tau_{m}(V)}\quad \frac{dh}{dt} =
\frac{h_{\infty}(V) - h }{\tau_{h}(V)}\\ \nonumber G_{K} &=&
\overline{G_{K}}n^{4} \quad \frac{dn}{dt} =
\frac{n_{\infty}(V) - n }{\tau_{n}(V)}\\
\end{eqnarray}
Here $m^{3}h,  n^{4}$ can be interpreted as the opening probability
of a channel. The $Na$ channel has two set of gates i.e., activation
gates represented by $m$ and inactivation gates represented by $h$.
The activation gates open and the inactivation gates close when the
membrane depolarizes. The $K$ channel has only single activation
variable which is a $4$ parameter system.

So we see that the state vector variables of the HH model are
$V,m,h,n$. The equations [1,2,3] can be written in a compact
matrix notation as
\begin{equation}\label{hh333}
\dot{\overrightarrow{X}} = \overrightarrow{F}(\overrightarrow{X})
\end{equation}
where $\overrightarrow{X} = [V,m,h,n]^{T}$. Equation [\ref{hh333}]
is a nonlinear equation and mainly numerical methods are employed
in solving such equations. We will not go into details of those
analysis as here we are interested to carry out the analysis in
terms of the relevance of quantum mechanics in ion channels
\cite{Vitiello95,Hameroff} and develop a framework for that and
then to see if there exist any limit for which it will reduce to
the HH model. In understanding the noise in ion channels apart
from the thermal background noise we can look into the phenomena
through the stochastic nature of closed(C) and open(O) states
with a distribution functions\cite{Umezawa},
\begin{equation}
\frac{O}{O+C}=\frac{1}{1+ exp[w_{0}-zeE]}
\end{equation}
where $w_{0}$ is the conformational energy in absence membrane
potential, z is the gating charge and E the membrane potential.
We speculate the possibility of stochasticity arising out of
closed and open channels. There has been an interesting result
related to the Runge Kutta numerical implementation of noise
terms \cite{fox,fox22} which are stochastic in nature associated
with some random numbers given by $$ \Delta\eta=\sqrt{{4\Delta t
log(r_{1}})cos(2\pi r_{2})}$$

More interestingly, the observation that understanding ion
channel dynamics is stochastic in nature has prompted us to look
at the relevance or analog of stochasticity in the quantum case. A
similarity of the HH model with the cellular automation has been
observed, \cite{fox,fox1,defil1} which in the limit of large ion
channel density gives rise to a Langevin description. Using the
Stratonovich description, the HH model is rewritten in the
Langevin form as

\begin{equation}
\frac{d}{dt}x_{i}= A_{i}(x)+ B_{ij}\eta_{j}(t)
\end{equation}
where $i,j = 1 \cdots n$ for the $n$ channels and $A_{i}, B_{ij}$
are related with the moments of the underlying transition
probability.

It is striking that HH formulation yields into a noisy model in
the large ion channel number limit. This observation has become
very crucial in our proposal of the general formulation of the HH
formalism in the quantum case. The most important theoretical
result we are hinting is the role of noise which is manifest in
the brain due to some functional geometry which is the underlying
structure assumed in the brain structure. The justification at
this stage will demand serious experimental investigations which
will explore the dependence of phenomena at various length scales.

We do also propose a new model based on the cooperative activation of sodium channels that reproduces the observed
dynamics of action potential initiation. In vitro experiments confirm this prediction, \cite{nat} supporting the
hypothesis that cooperative sodium channel activation underlies the dynamics of action potential initiation in cortical neurons.
The rapid onset of action potentials is independent of the temporal structure of synaptic inputs and of the
 electrophysiological cell class. It seems that rapid action potential onset and large variability in onset
 potentials are strongly antagonistic in Hodgkin–Huxley-type models. In such models, the initial phase of an action potential
 is determined by the activation of voltage-dependent sodium channels. Their dynamics is described by the activation curve and
 kinetics of an associated gating variable. In the Hodgkin–Huxley formulation it can be shown that the rate of membrane
 depolarization is limited by $g_{Na}h m^{3}(V)(V_{Na} - V)/C + I_{o}/C$, where $g_{Na}$ denotes peak sodium conductance,
 $h$ is the fraction of sodium channels available for activation, $m^3(V)$ is their activation curve, $V_{Na}$ is the sodium
  reversal potential, C the membrane capacitance, and $I_{o}$ is the current carried by other channels.
  It is plausible that there is a one-to-one relationship between the single-channel activation curve and the action
  potential onset dynamics, owing to the assumption that the opening of individual sodium channels is statistically independent.
  This assumption, however, might be violated in the highly organized molecular machinery of a living cell \cite{smart}.
   Indeed, the rapid onset of action potentials suggests that many sodium channels open virtually simultaneously, that is,
   in a potentially cooperative fashion.

To assess whether cooperative activation of voltage-gated sodium channels can account for the two characteristic features of cortical action potential initiation, a model of a population of coupled sodium channels was constructed. Assuming that channel interactions are distance-dependent in neuronal membranes, the model predicts  gradual recovery of the number of available sodium channels during washout of TTX led to a gradual increase in action potential onset rapidness. These results cannot be explained by Hodgkin–Huxley-type models, in which reduction in the sodium channel density modifies only the amplitude of action potentials and their onset potential, but not their onset rapidness.

\section{Nature of Brain Processes}
It is really debatable at this stage to pin the nature of brain processes. Inspite of the minuteness of the ion channels or the neurons, it really seems that decoherence effects will nullify the quantum effects of the neurons as a whole. But in the light of recent discoveries on , cooperative phenomena and some stochastic dynamics in ion channels, it is really true that even quantum effects in single ion channels, will cumulatively give rise to some hitherto unknown classical states. As an example we have summarized decoherence processes in ion channels due to some fundamental processes.
\def\filler{\hbox{$\quad\quad$}}
\begin{center}
{\footnotesize
{\bf } Decoherence timescales.
\noindent
\begin{tabular}{llc}
\noalign{\vskip 4pt}
\hline
\hline
Object      & Environment       & $t_{dec}$\\
\hline
\noalign{\vskip 2pt}
Neuron      & Colliding ion     &$10^{-20}$s\\
Neuron      & Colliding $water$\filler&$10^{-20}$s\\
Neuron      & Nearby ion        &$10^{-19}$s\\
Microtubule\filler &Distant ion     &$10^{-13}$s\\
\hline
\hline
\end{tabular}
}
\end{center}
The results may enable us to address the question of whether
cognitive processes in the brain constitute
a classical or quantum system.
Neuron firing itself is also highly classical, since it occurs
on a timescale $t_{dyn}\sim 10^{-3}-10^{-4}\,seconds$
\cite{Ritchie}.But the problem with such approach is closely related to assuming that there is a unique timescale associated.
For example if we admit a quantum description inside the ion channels then, because of the uncertainities introduced at the
 ionic level, the brain state will develop into a continuos distribution of virtual macroscopic states.

\section{\bf Stochastic geometry in
Neuronal Modeling }\label{secstoch}

Very Recently the (Nonlinear Schrodinger Equation) NLSE has been
solved with an artificial neural network scheme. This analysis
gives us an insight and assurance  that maybe the NLSE will play
an important role in analyzing realistic neuronal modeling. Here
we discuss in brief about the solution of NLSE on a network.

The time dependent propagation of light pulse inside a single mode
nonlinear optical fiber is given by the solution of
\begin{equation}
i(\frac{\partial \Psi}{\partial z}) -\alpha (\frac{\partial^{2}
\Psi}{\partial z ^{2}}) - \beta {\Vert{\Psi}\Vert}^{2}\Psi = 0
\end{equation}
where $\Psi$ is the field amplitude, $z$ and $t$ are the optical and
time axis respectively, $\alpha, \beta$ are the dispersive and
waveguide coefficients respectively. The competition between pulse
dispersion and focussing gives rise to the formation of solitons for
a particular input. With suitable boundary conditions a stable
soliton is obtained. It has been observed that the solution consists
of a 3 layer architecture with 42 hidden nodes \cite{nn1}. Now to
speak of the implications of this result it has been also observed
that the knowledge of the upper bound on the field amplitude
provides a stopping criterion on the training of the neural network
(NN).

We present a model where the propagation of activity is
stochastic and the connections are random. Each excitable element
$i = 1,\ldots, N$ has $n$ states: $s_i = 0$ is the resting state,
$s_i = 1$ corresponds to excitation and the remaining $s_i =
2,\ldots, n-1$ are refractory states. There are two ways for the
$i$-th element to go from state $s_i = 0$ to $s_i = 1$: a) due to
an external signal, modelled here by a Poisson process with rate
$r$ (which implies a transition with probability
$\lambda=1-\exp(-r\Delta t)$ per time step); b) with probability
$p_{ij}$, due to a neighbour $j$ being in the excited state in
the previous time step. Time is discrete (we assume $\Delta t =
1$ ms) and the dynamics, after excitation, is deterministic: if
$s_i = 1$, then in the next time step its state changes to $s_i
=2$ and so on until the state $s_i = n-1$ leads to the $s_i = 0$
resting state, so the element is a cyclic cellular automaton,
\cite{Marro99,olfac,adams}. The Poisson rate $r$ will be assumed
to be proportional to the stimulus level $S$ .

Two kinds of oscillations are observed in the system. Under
sufficiently strong stimulation, {\em all\/} networks present
transient collective oscillations, with frequencies of the order
of the inverse refractory period. They are a simple consequence
of the excitable dynamics and the sudden synchronous activation by
stimulus initiation.  The transient behaviour is reminiscent of
oscillations widely observed in experiments\cite{Laurent02}, but
its trivial origin suggests that they are epiphenomenal and
without computational relevance. Networks with $\sigma>
\sigma_{osc} > \sigma_c$ also present self-sustained oscillations
in the absence of stimulus where $\sigma_{osc}$ is a bifurcation
threshold. The frequency depends on the network parameters, but
remain in the gamma range. The oscillations are similar to
reentrant activity found in other models of electrically coupled
networks\cite{Lewis01}.

It may be pertinent to ask at this stage that what use is of the
above scheme to our proposed model. What we believe is that
applicability of NLSE on NN gives us a clue that may be the
Schr\"odinger equation is applicable at diverse length \&\ temporal
scales in neuronal architecture with an unknown, a priori geometry
and the basic objective is to find out the appropriate dynamics
for that.

We have already emphasized that the neuronal architecture has a form
of geometry with some probabilistic structure on it giving rise to a
probabilistic manifold \cite{prob}. So the main point of the
analysis depends on the identification of a stochastic
interpretation to quantum mechanics. The essential ingredient is
following. We claim that the
 operator $A = b_{\nu}(x)\partial_{\nu} + (\frac{\hbar}{2\pi i}) {\nabla} $ is the
  infinitesimal  generator of the stochastic process defined by the Langevin equation
\begin{equation}
dx_{\mu}(t) = b_{\mu}(x,t) dt + dW_{\mu}(t)
\end{equation}
The importance of this identification is that classical probability
theory gets related with quantum mechanics. Now, the next question
what we can ask is that we are trying to define the SE in a curved
probabilistic manifold. So, apart from a stochastic approach to
quantum mechanics ,we need something more, i.e., to randomize the
metric. Let us assume a Lagrangian, given by

\begin{equation}
L = \frac{m}{2}g_{\mu\nu}(x) {\dot{x}}_{\mu}{\dot{x}}_{\nu} -V(x)
\end{equation}

 Variation of this Lagrangian gives us the equation of motion in the form of geodesics.
  If we vary the trajectories and define a stochastic process in terms of the variations
  with gaussian spread and compare this distribution with the Feynman path integral, we end
  up with the Riccati equation which is the stochastic analogue of the Schr\"odinger equation
   on a curved manifold.
 \begin{equation}
 \label{stoch1}
\frac{\hbar}{2}{\nabla}_{\mu} x^{\mu} + \frac{m}{2}x^{\mu}x_{\mu} =
V + \frac{{\hbar}^2R}{6m}
\end{equation}

So we see that stochasticity \cite{stoch00} involves a generation
of an effective potential of motion. We will see now that how this
may be handled in analyzing the quantum circuits.

\section{Rules of Design through Cooperativity \&\ Quantum Mechanics }
It is indeed true that in understanding neural mechanisms, we need
nonlinear and dissipative analysis. Now as has been argued in
\cite{qmic3} over the years and until recently, if we think of the
relevance of quantum mechanics and the role of Cooperative
phenomena for neuronal dynamics at suitable length scale. For an
atom in a cavity, a process such as spontaneous emission is
sometimes viewed as dissipative but if some number of modes are
chopped of, the process becomes reversible. In comparison, the
resistance to electric current flow is reversible, which is
typical of closed system. But if we think of quantum circuits the
situation is drastically different.

In this context, we cite an particular example: A current driven RC
circuit which is identical to a free particle driven by an external
force. In absence of the resistor, the system is well described by
the charge operator $Q$ and operator $\phi$ which satisfy $ [\phi,
Q] = i\hbar$. We would like to mention here that a circuit theory
\cite{ckt00,ckt} that can describe quantum transport, is
particularly important and has potential applications in
nanotechnology, molecular devices and beam epitaxy etc \cite{nano}.

As we have already seen that the electric charges of ions are in
fact responsible for the membrane potential and action potential.
Generation of the potential therefore gives rise to the possibility
of modeling the ion channels through electric circuits, which
generate the required potentials. The scheme is devised through a
quantum analogue of the corresponding electrical circuit. So our
objective is pretty clear. Some very recent results at ionic scales
regarding the relevance of Quantum Mechanics (QM) \cite{qmic4}, we try to build
some viable neuronal models and corresponding electric circuits \cite{ckt3}. But
as QM governs the dominant dynamics we have to develop quantum
circuits.

In this context, we would like to mention very important work
\cite{qcio2} which we earlier mentioned, related to the circuit
equivalent of Schrodinger's equation. The circuits were originally
designed for completely different purposes and had no connection
with brain activity. It is a way to measure the eigenvalues,
eigenvectors and statistical means of various operators, belonging
to the system which are being modeled by electrical means. We
briefly discus below the scheme for handling those things.

Let the wave equation be divided by $i\omega_{c}$ where $\omega_{c}
= \sqrt{\omega} = (E/\hbar)^{\frac{1}{2}}$ and multiplied by
$\triangle x$ we get
\begin{equation}
-\frac{1}{\omega_{c}}\frac{\hbar^{2}}{2m}\frac{1}{\triangle
x}\frac{\partial^{2}\psi}{\partial x^{2}}\triangle x^{2} + \Big(
\frac{V\triangle x}{i\omega_{c}} + \hbar i \omega_{c}\Big) \psi = 0
\end{equation}

We should like to note some salient points here.
 \begin{itemize}
\item The Kinetic energy operator $T$ is represented by a set of
inductors in series whose inductance is given by $ L_{1} =
\frac{2m}{\hbar ^2}\triangle x$.

\item The potential energy operator
V is represented by a set of unequal coils in parallel whose
inductance is $L_{2}= 1/V\triangle x$.

\item The total energy
operator $- E$ is represented by a set of equal capacitors whose
capacitance is $\hbar\triangle x$.

\item The operand $\psi$ is
represented by voltages and the result of the operation $\alpha
\psi$ where $ \alpha$ is any operator, is represented by currents.
 \end {itemize}

This model can also be extended to nonorthogonal coordinates and in
general on arbitrary manifolds. Utilizing the circuits, tests were
carried out on an ac network analyzer. The results are worth
mentioning. The tests were done in 1-dimensional circuits. or
example measurements were made for a particular case of the
rectangular potential well and analyzed which had good agreement
with the experimental results \cite{exp}.

For the sake of completeness, we should like to mention here that
from the preceding model, we can develop a prescription for
developing a electric circuit equivalent for the Schr\"odinger
equation (SE). The SE has some analogies with the heat conduction
equation

\begin{equation}
\frac{d\psi}{dt} = \frac{1}{\hbar}\Big(\frac{\hbar^{2}}{2m}
\nabla^{2} + V \Big)\psi
\end{equation}

Now we make a prescription for the electric circuit as equivalent to
the Schr\"odinger equation as

\begin{eqnarray}
V\Leftrightarrow \psi \quad \frac{1}{r}\Leftrightarrow V\triangle
Vol \quad \frac{1}{R x}\Leftrightarrow \frac{\hbar^{2}}{2m}
\triangle Vol/(\triangle x^{2}) \quad C\Leftrightarrow i\hbar
\triangle Vol
\end{eqnarray}

The construction relies here on having $N$ imaginary capacitance and
one of the consequence is we will get solutions of the form

$$ \psi\sim \exp(ikx) \exp(i\omega t)$$.
Physically this means that we don't
have exponential decay with time into thermal equilibrium but we get
everlasting solutions which conserve $|\psi|^{2}$. So we see that if we impose ccoperativity into the picture using theory
of resonances, and by dint of central limit theorem, it is plausible to get the required behaviour of non solitonic
behaviour corresponding to the fields.

\section {Quantum Circuits at Nano Length Scales}

Motivated by the quantum mechanical considerations and the circuit
equivalence of Schr\"odinger equation we will now try to formulate
an equivalent circuit for the membrane potential in the ionic
channels. We will consider a single ion channel and consider the
circuit implementation for it. But there are some subtleties
regarding this. Tensor network theory \cite{tnet} may be realized in
the brain and there is possibility  of a non trivial geometrical
structure inside brain as mentioned in \cite{ama12,qmic5,prib}, work also points towards
this direction. This implies that there is an underlying geometrical
structure inside brain and it may be important at the ionic scales.
So we try to develop a formalism for describing that. We start with
some simplifying propositions:
\begin{itemize}

\item  There is an underlying geometry inside brain which is
responsible for neuronal activities and the geometry can be
described by a metric.
\item  Quantum mechanics is applicable at the length scales of ionic
channels and the phenomena can be described by Schr\"odinger
equation
\item  The phenomena at those length scales is stochastic.
\end{itemize}

With these propositions we can now think of a formalism for
various set of events inside the brain. It should be mentioned
here that ultimately we would like to connect our formalism to
the HH formalism, which has been successful in describing the
membrane gating and dynamical phenomena involving channels
\cite{ic}.

So we start by writing a SE equation on a curved manifold. The
equation can be identified with a Heat Kernel equation.
\begin{equation}
i\hbar\frac{\partial \psi}{ \partial t} =
\frac{1}{\sqrt{g}}\partial_{\mu}(\sqrt{g}\partial_{\mu}\psi) + V\psi
\end{equation}
The equation above, for arbitrary metric, is in general, nonlinear
and in accordance with our third proposition we claim that the
processes are stochastic in nature and the analysis of section
\ref{secstoch} tells us that the dynamics will be governed by
Riccati like equation [\ref{stoch1}] with a correction in the
potential term as

\begin{equation}
i\hbar\frac{\partial \psi}{ \partial t} = \frac{\hbar^{2}}{2m}
\nabla^{2}\psi + (V + \frac{\chi (q)
\hbar^{2}\sqrt{N}\vartheta}{6})\psi
\end{equation}

$\vartheta$ is the curvature scalar, associated with the metric,
which gets incorporated into the potential term. It is important to
note that the correction term differs from that of equation
[\ref{stoch1}] in a coordinate dependent factor. We will show that
this factor is crucial in developing a quantum equivalent of a
circuit. The mass is absorbed in the coordinate dependent term. The $\chi$ in some sense acts as a space
modulation factor which governs both the noncommutative aspects and the fluctuations from the metric.It
should be mentioned here that this equation describes the dynamics
for $N$ channels and we have made a conjecture by including the
number of channels, with a hope to get the classical picture of the
HH equation in global limit.

The Schr\"odinger equation along with the correction of the quantum
term is a good starting point in our case to develop a circuit
equivalent. Actually the problem in this case to extend the previous
construction for the equivalence of SE to electrical circuit is
dictated by the presence of the metric. First of all, the metric is
a dynamical variable which governs the behavior of space time. So,
we cannot just implement it as some electrical component, since such
a component should have the ability to shape the global structure of
the full circuit. At the moment, we do not know of any such
component. We have seen that in periodically driven circuits, we can
scale the capacitance or say the inductance as $ L \sim \gamma g(t)
$ which may capture, in some sense, the global dynamics, but it will
not catch the full glimpse of the dynamical behavior. The curvature
gets into the potential term and thereby fluctuations in the metric
will induce different potentials and hence only periodic variations
may not do \cite{sr11,sr22}. It is a widely held that fundamental
processes of nature may be explained by probabilistic metric and the
probabilistic features can be modeled into uncertainties or
fluctuations from a physical point of view. If we introduce the
fluctuations in the metric as
$$ g_{ij}(x, h) = g_{ij}(x) + \alpha_{ij}(h) $$
 the fluctuation of the
metric generates a random potential $ V$, a random coefficient $S$
which depends on the fluctuations. In the quantum case we will do
indeed get dissipation which depends on the fluctuation. The
Schr\"odinger equation turns to be
\begin{equation}
\frac{\hbar^{2}}{2} \frac{\partial^2 \phi}{\partial r^{2}} + V\phi =
 S\frac{\partial \phi}{\partial t}
\end{equation}

To make connections with brain activities and neuronal circuits we
try to develop circuits corresponding to quantum mechanics, the
circuit will do contain some flavor of the noncommutative aspects.

We know that brain phenomena is considered as dissipative. In such
kind of theories, one considers such one partcle dissipation in
quantum theory. So, we try to extend that formalism in our case with
the corrected potential along with a source term. Then we consider
the following Hamiltonian as:
\begin{equation}
H = -e^{-Rt/L}\frac{\hbar^{2}}{2e^{2}L}p^{2} +
e^{Rt/L}(\frac{1}{2C}q^{2} + \varepsilon q +
\frac{\chi(q)\vartheta \hbar^{2}\sqrt{N}}{6})
\end{equation}
Here $q$ is state variable, $p$ the conjugate which goes uplifted to
the charge operator when we deal with quantum mechanics (QM). Using
the Heisenberg equation of motion, the equation for the state
variable is given by
\begin{equation}
L\frac{d^{2}q}{dt^{2}} = \frac{1}{2e^{2}}(\{\frac{1}{2C}q^{2}
+\varepsilon q + \frac{\chi(q) \vartheta \hbar^{2}\sqrt{N}}{6}\},
[p^{2}, q]) - R\dot{q}
\end{equation}
So, evaluation of the simple commutator gives us the equation for
the corresponding quantum circuit as
\begin{equation} L\ddot{q} +
R\dot{q} + \frac{q}{C} = \varepsilon +
\frac{\alpha\hbar^{2}\sqrt{N}}{6}\vartheta \dot{q} \int {dt
\chi'(q)}
\end{equation}
The last term  in the above equation is most striking. It shows that
that the above equation clearly shows that the inductance gets
corrected, by a quantum term. In this way, ultimately, at the level
of circuit equivalence, we will be getting a renormalized
inductance. It gives the equation a status of an integral equation
and would be interesting to find out the conditions under which it
will reduce to a differential equation. In that case, the
capacitance gets renormalized.

It is very important so as to make some measurements to find out
these extra factors. There is an extra parameter in the theory
which needs to be fine tuned to get the desired effects. The above
equation can also be transformed into a Langevin like form and
get a measure for the Probability functional, which is very non
trivial due to the presence of the curvature term and may hint at
some statistical manifold like character. This is not quite
surprising as we mentioned at the beginning of the section that
it may arise due to the intrinsic stochasticity of the neuronal
activities. The above observations have some interesting
consequences with respect to Nonlinear Schrodinger Equation. Some
of the results is worth mentioning . If we had included in the
Schr\"odinger equation a damping factor in the form of a bounded
negative operator and a quasi periodic force, the solutions turn
out to be even and periodic. The analysis in such case, give rise
to the existence of invariant manifolds in the phase space of the
equation. The infinitely many eigenvalues in the integrable limit
turn into complex eigenvalues with negative real parts. The
manifolds exhibit a dynamical behavior and the geometry resembles
those of certain homolinic orbits in finite dimensional Ordinary
Differential Equation \cite{slink}.

\section{Determinism as a limit of Underlying Stochastic Processes}

It is now pertinent to understand how the stochastic quantum phenomena at ionic scales by decohorence effects give rise to determinism of the HH models.
The instantaneous electrical state of the
axon depends on the locations and internal
states of any molecular mechanisms at work in the axon and in particular ion channels.As we have mentioned before the stochastic model
describes the working of individual ion channels, whereas the
deterministic model `averages out' their behaviour, involving
instead functions that describe the proportion of those channels in
a small neighbourhood of a point that are in each possible state.

The state of our system is partly described by a function $v:I
\rightarrow\textbf{R}$ giving the value of the membrane
potential at each point along the axon. Since ions can diffuse
along the axon, the variation of this function with time will
also exhibit diffusive behaviour, allowing us to impose certain
regularity conditions on it. Our model involves a renormalization
of ionic conductivities corresponging to each ion channel pore.
Assuming all channels as identical, as described above, in our
model the driving potentials  will actually correspond to the
different possible channel states $\xi \in E$, the space of
states. The stochastic model implies that a channel at position
$x$ will jump between states $\xi,\zeta$ at random at specific
rates $\chi_{\xi,\zeta}(V)$, where $V$ is the value of the
potential difference at the relevant point $x$. So there will be
one such model for each $N \in \textbf{N}$. The deterministic
model arises heuristically as the limit of the stochastic model
with very many very small ion channels; that is, for large $N$.
In the deterministic model a new family of functions $p_\xi \in
\rm{Lip}(I,[0,1])$ for $\xi \in E$, may be introduced that
replicates the role of the individual-channel configurations. The
value $p_\xi(x)$ is to be interpreted as the proportion of those
channels in a small neighbourhood of the point $x$ that are in
state $\xi$.

\section{\bf Conclusion}

The primary observable of an ion channel is its conductance. Because the channel current depends on both applied voltage (V) and bath concentrations (C) and the conductance data is typically plotted in the form I-V \&\ I-C Curves. We need to estimate in our case the conductivities for the ion channels from the curves, which follows non-linearity, upto an appreciable range. Simulation model studies have indicated that the non-linearities arise as a result of residual energy barrier in the channel. A Quantum Jump approach analogy for understanding how the dynamics of Stochastic Schr\"{o}dinger Differential Equation induces cooperative mechanism using the damping of one field mode in a cavity at temperature T can also be given.

The striking aspect of our result is that in the most general case,
for scales in which QM is applicable, we have found out a
generalized HH equation with the conductances $ G_{A}$ being
corrected with the renormalized value in equation [\ref{hh111}] by

$$ G_{A}+ ({\frac{\alpha\hbar^{2}\sqrt{N}}{6}\vartheta \dot{q}\int {dt
\chi'(q)}})^{-1}$$

It is necessary to study following two issues. Firstly, to see
under what limit does this modified equation i.e., generalized HH
equation turns to ordinary equation with no renormalization.
Then, one needs to do the experiments to see whether the
conductances indeed do get corrected. If it is so then  we could
measure such term for single ion channels. We also believe that
one of the mechanisms by which we may get ordinary HH theory with
no renormalization (i.e quantum mechanics is unimportant) is when
there are many channels and quantum mechanics is getting subdued
in the large $N$ limit. Anyway, there is a subtle point here. In
confirmation of the relevant observation for the stochasticity of
HH equation in the Langevin description, we see that in the
classical limit, we may get stochasticity for a critical large
value of the number of channels. It is really important to design
experiments to measure critical parameters as appeared in the
above equation. Such experiments will be very conclusive for the
correctness of the model and also give a direct evidence for the
applicability of QM in ion channels. It is also crucial to
measure the effective conductance. At this stage, we still do not
know how we can model such mechanisms, but experimental results
on single ion channel may surely shed some light in understanding
these aspects \cite{exp00,exp001,montal,sankar}.

\section*{Acknowledgement}
IM \&\ GH wishes to thank B \&\ B funding at Georgia State University (GSU) which helped immensely for the completion of this work. IM also ackowledges the hospitality of SR at George Mason University where some crucial ideas reagrding this paper evolved. SR is thankful to College of Science, GMU for support during a crucial part of this work.

\end{document}